\documentclass{nato}  

\usepackage{txfonts}

\usepackage[draft]{hyperref}

\newdisplay{guess}{Conjecture}

\begin{document}


\begin{opening}
\title{New proposal of numerical modelling of Bose-Einstein
correlations: ~ Bose-Einstein correlations from "within"\thanks{Invited
talk delivered by G.Wilk at the International Conference {\it NEW TRENDS
IN HIGH-ENERGY PHYSICS (experiment, phenomenology, theory)}, Yalta,
Crimea, Ukraine, September 10-17, 2005. }}

\author{Oleg V. Utyuzh \email{utyuzh@fuw.edu.pl}}
\institute{The Andrzej So\l tan Institute for Nuclear Studies;
Ho\.za 69; 00-681 Warsaw, Poland }

\author{Grzegorz Wilk \email{wilk@fuw.edu.pl}}
\runningauthor{G. Wilk} \runningtitle{Bose-Einstein correlations from
"within"} \institute{The Andrzej So\l tan Institute for Nuclear Studies;
Ho\.za 69; 00-681 Warsaw, Poland}

\author{Zbigniew W\l odarczyk \email{wlod@pu.kielce.pl}}
\institute{Institute of Physics, \'Swi\c{e}tokrzyska
Academy, \'Swi\c{e}tokrzyska 15; 25-406 Kielce, Poland}

\begin{abstract}
We describe an attempt to numerically model Bose-Einstein correlations
(BEC) from "within", i.e., by using them as the most fundamental
ingredient of  a Monte Carlo event generator (MC) rather than considering
them as a kind of (more or less important, depending on the actual
situation) "afterburner", which inevitably changes the original physical
content of the MC code used to model multiparticle production process.
\end{abstract}

\end{opening}

\section{Introduction}

In all multiparticle production processes one observes specific
correlations caused by quantum statistics satisfied by the produced
secondaries. Because majority of them are boson (pions, kaons,...)
subjected to Bose-Einstein statistics, one usually observes enhancements
in the yields of pairs of identical boson produced with small relative
momenta. They are called Bose-Einstein correlations \cite{BEC,TC,WK,GA}.
The corresponding depletion observed for fermionic identical particles
(like nucleons or lambdas) satisfying Fermi-Dirac statistics is also
observed but we shall not discuss it here. There is already extraordinary
vast literature ( see, for example, \cite{TC,WK,GA}, and references
therein), to which we refer, in what concerns measurements, formulations
and interpretation of BEC as well as expectations they cause for our
understanding of spatio-temporal details of the multiparticle production
processes (or hadronization processes). Here we shall concentrate only on
one particular field, namely on the numerical modelling of BEC and we
shall propose new approach to this problem, which we call {\it BEC from
within} \cite{QCM}.

The need for numerical modelling of the BEC phenomenon arises because the
only effective way to investigate multiparticle production processes is
by using one out of many of Monte Carlo (MC) numerical codes developed so
far, each based on different physical picture of hadronization process
\cite{MC}. However, such codes do not contain BEC because they are based
on classical probabilistic schemes whereas BEC is of the purely quantum
mechanical origin. Modelling BEC is therefore equivalent to modelling
quantum mechanical part of hadronization process and by definition can
only be some, better or worse, approximation. Actually, so far there is
only one attempt in the literature of proper modelling of BEC using them
as input of the numerical code. Namely in \cite{OMT}, using information
theory approach based on the maximalization of the information (Shannon)
entropy, identical particles were allocated in separated cells in the
(longitudinal in this case) momentum space. In this way one is at the
same time obtaining proper multiplicity distributions and bunching of
particles in momentum space mimicking bunching of bosons in energy
states, i.e., effect of the Bose-Einstein statistics. The method we shall
propose here develops this idea further. However, before continuing along
these lines let us first mention other attempts of introducing effects of
BEC, which are widely used nowadays and let us point out why they cannot
be regarded as satisfactory.

\section{Numerical modelling of BEC}

Among methods of numerical modelling of BEC using, in one or another way,
some known numerical MC codes one can distinguish two approaches. One,
represented by \cite{ZAJC} (working with plane waves) and \cite{MP}
(working with wave packets) starts with some distribution of momenta of
particles in selected event and then, using a kind of Metropolis
rejection method, changes step-by-step their momenta until multiparticle
distribution containing effect of BEC is reached. This is now regarded as
a true event. The drawback of this method is that it is extremely time
consuming and therefore can only be used to explain some fine details of
how BEC works, not for comparison with real data.

The other approaches are tacitly assuming that, on the whole, BEC
constitutes only a small effect and it is therefore justify to add it in
some way to the already known outputs of the MC event generators widely
used to model results of high energy collisions in the form of the so
called {\it afterburner} \cite{After}. There are two types of such
afterburners.
\begin{itemize}
\item The first one modifies accordingly energy-momenta of identical
secondaries, \textbf{p} $\rightarrow$ \textbf{p} + d\textbf{p}, where
d\textbf{p} are selected from some function $f({\rm d}$\textbf{p}$)$
chosen in such way as to reproduce the observed correlation pattern
\cite{Sj}. The advantage of this approach is that it applies to a single
event and does not change multiplicity of secondaries produced in this
event. The obvious drawback is that it spoils energy-momentum
conservation so afterwards one has to correct for it in some way.

\item The second type of afterburners preserves the energy-momentum
conservation but it changes multiplicity distribution pattern of given MC
code and works only for the whole ensemble of events produced by MC. In
it one weights each event depending on how big BEC effect it shows
(because of inevitably fluctuations they can be events which already show
quite substantial BEC effect - those are multiplied by bigger weight -
together with events which show no BEC - they are multiplied by small
weight). In practice it leads to distortion of physics because "suitable"
events are counted many times more than the MC code used allows
\cite{Fj,And}. Again, the weight function is some phenomenological input
chosen in such way as to reproduce the observed BEC pattern of the whole
ensemble of events considered.
\end{itemize}

\begin{figure}
\footnotesize \centerline{\begin{tabular}{@{}cc@{}}
\includegraphics[height=1.4in]{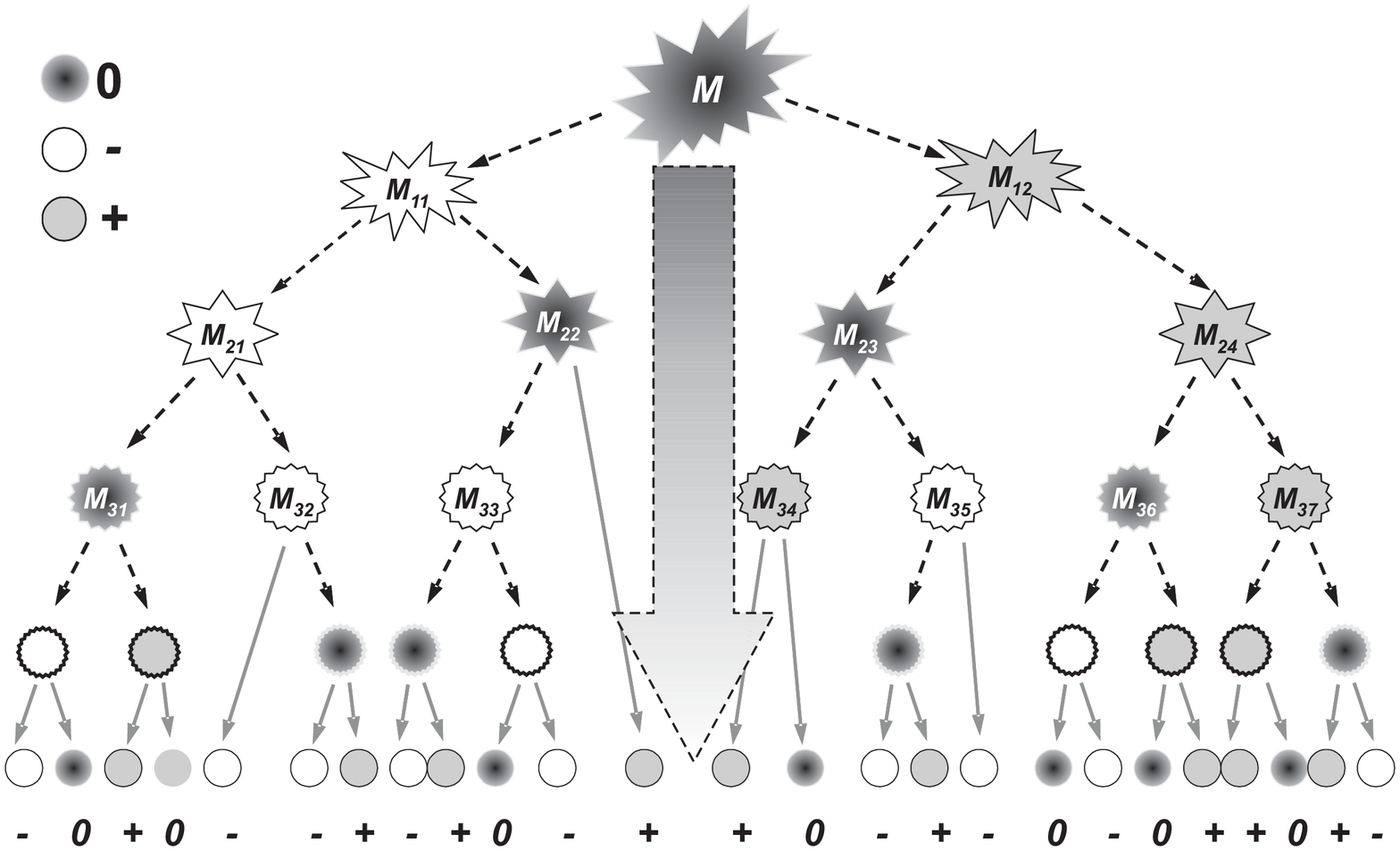} &
\includegraphics[height=1.4in]{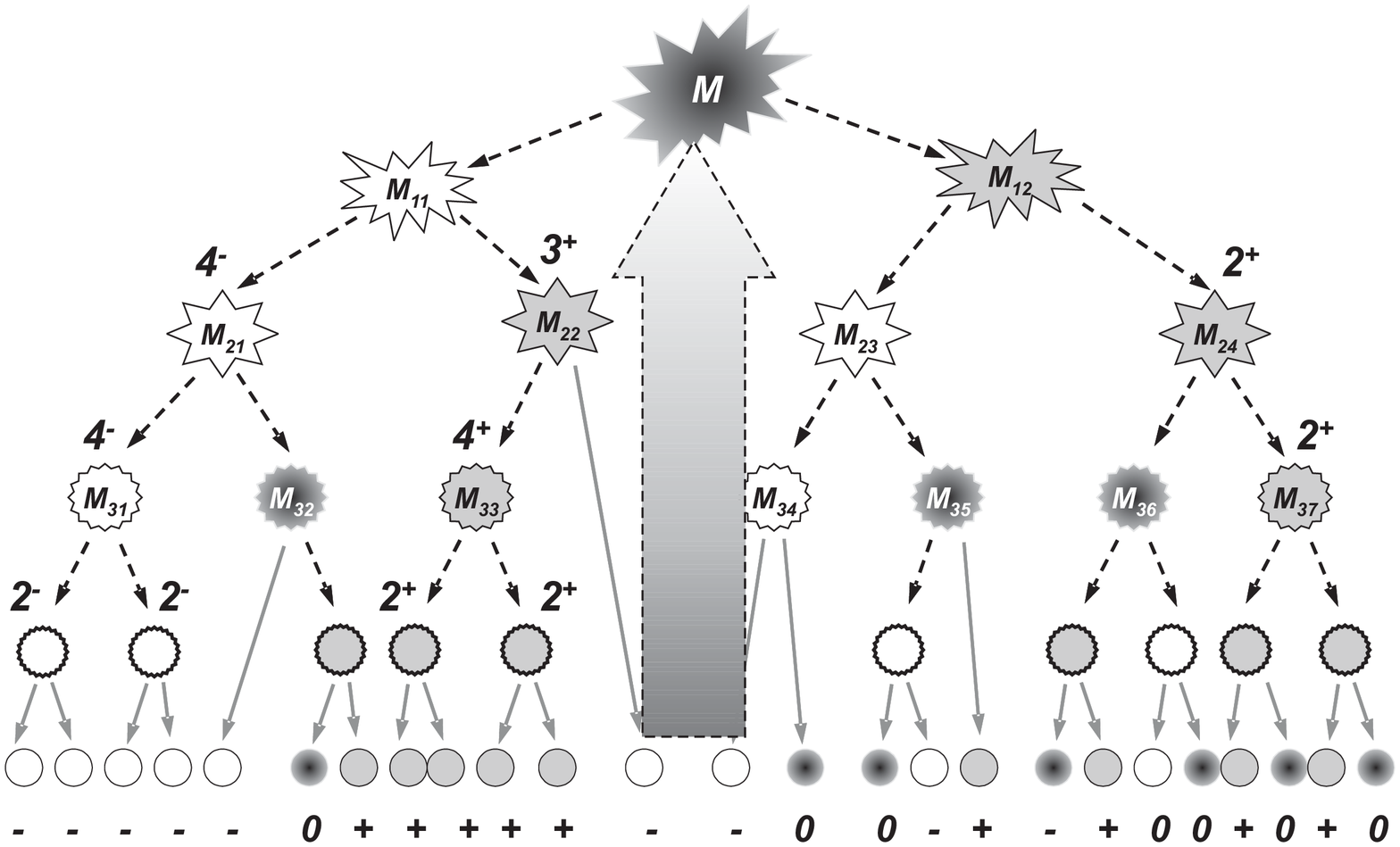}\\
\end{tabular}}
\caption[]{Example of charge flows in MC code using simple cascade model
           for hadronization \protect\cite{CAS}: left panel - no effect of BEC
           observed; right panel - after applying afterburner described in
           \protect\cite{UWW} (based on new assignement of charges to the produced
           particles) one has BEC present at the cost of appearance
           of multicharged vertices.}\label{Flow}
\end{figure}

\begin{figure}
\footnotesize \centerline{\begin{tabular}{@{}ccc@{}}
\includegraphics[height=2.2in]{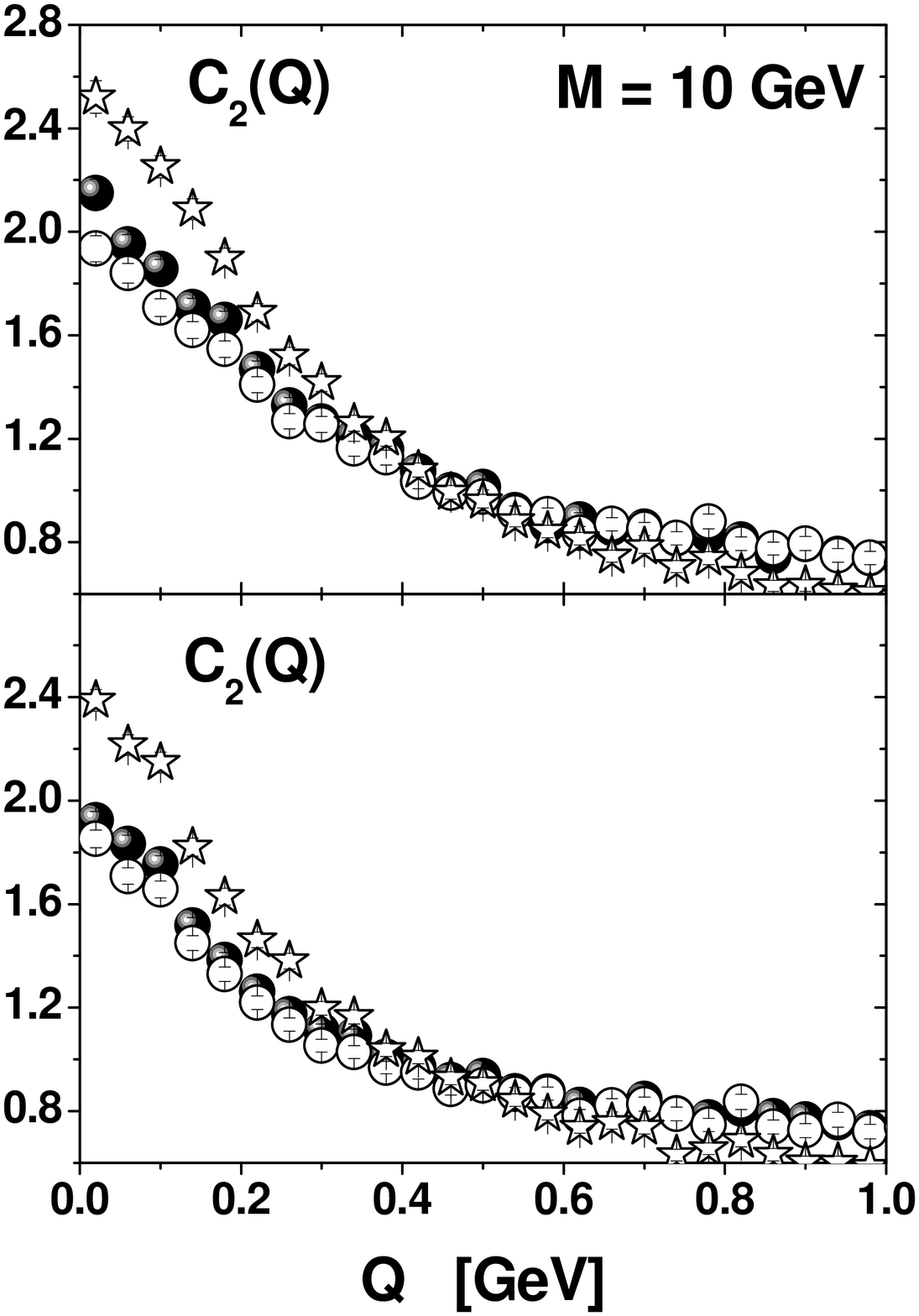} &
\includegraphics[height=2.2in]{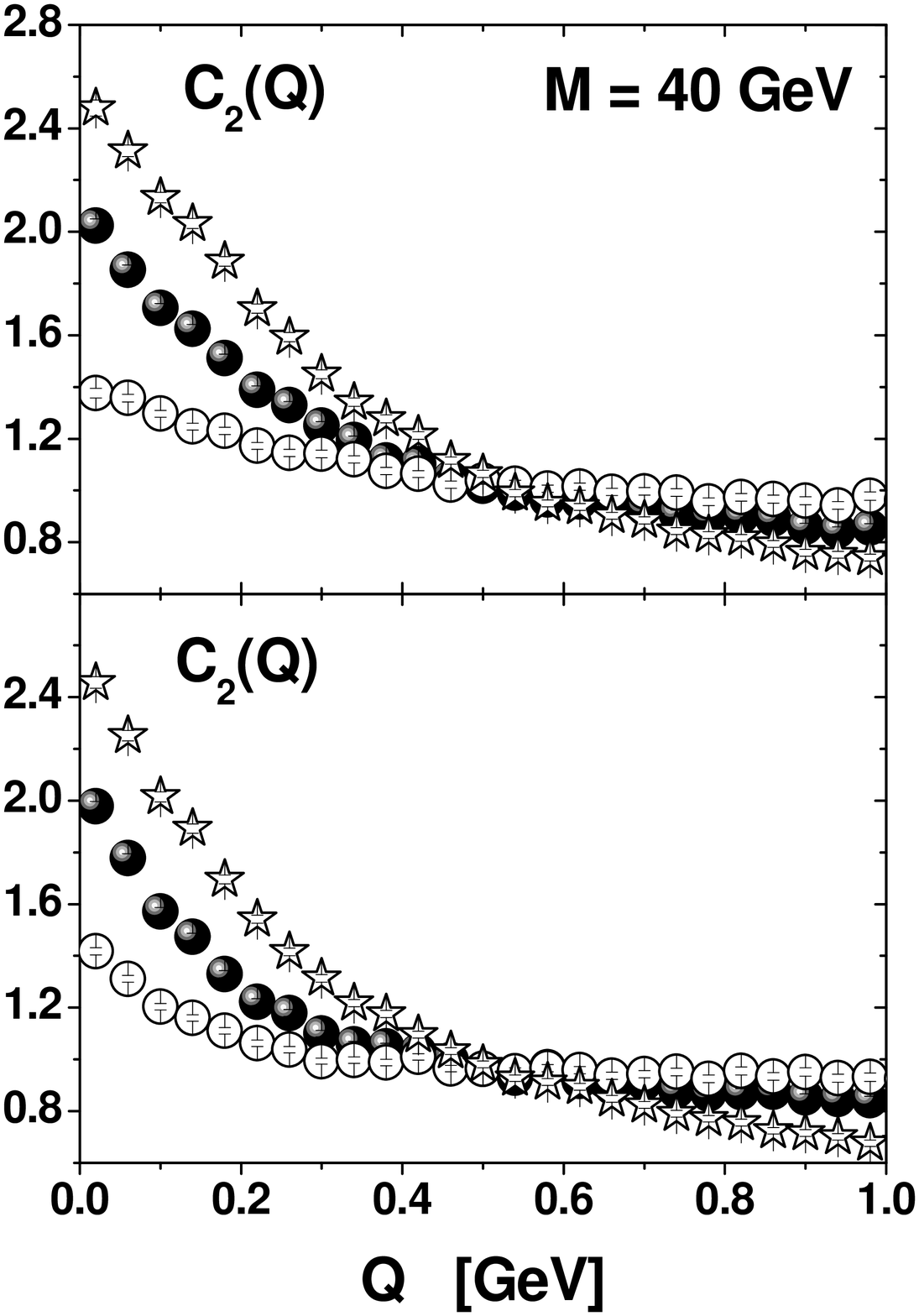} &
\includegraphics[height=2.2in]{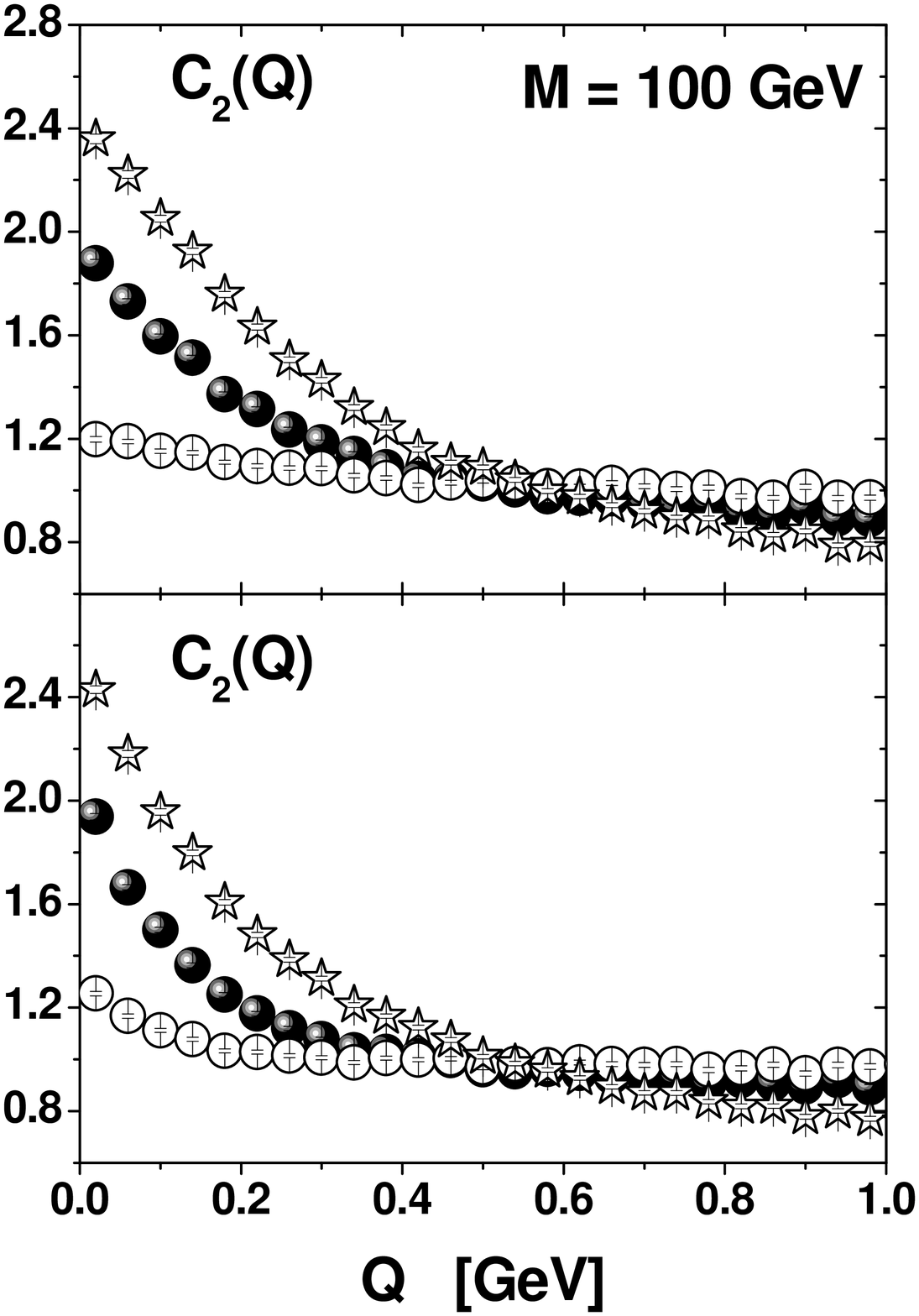}\\
\end{tabular}}
\caption[]{Examples of BEC patterns obtained for $M=10$, $40$ and $100$
GeV for constant weights $P=0.75$ (stars) and $P=0.5$ (full symbols) and
for the weight given by gaussian weights build on the information on
momenta and positions of particles considered \cite{SP} (open symbols).
Upper panels are for CAS, lower for simple statistical model (see
\cite{SP} for details).}\label{C2M}
\end{figure}

However, even - as it is usually assumed - if the above afterburners do
not change in a substantial way the {\it numerical} output of MC
generators, which they are using,  they surely do change their {\it
physical basis} (i.e., the number and the type of the initial physical
assumptions forming a basis of such MC code). This change, neglected in
the assumption made above, has never been investigated in detail. We
shall now demonstrate how dramatic this change can be by using as example
simple MC cascade code (CAS) \cite{CAS} proposed by us some time ago and
endow it with some specific afterburner. The physical basis of CAS is
very simple. It starts with decay of the original mass $M$ into two
objects of mass $M_1$ and $M_2$ such that $M_{i=1,2} = k_{i=1,2}\cdot M$
where $k_{i=1,2} <1$ are parameters responsible for development of the
cascade ($k_1 + k_2 <1$). The masses $M_{i=1,2}$ fly then apart and after
some time $t_{i=1,2}$ they decay (in their rest frame) again into two
lighter masses according to the above scheme but now with different
values of the new decay factors $k$. This branching process continues
until the original mass $M$ finally dissipates into a number of masses
$M_{i=1,\dots,2^N}$ equal to masses of the lightest particles ($N$ is the
number of cascade generations). Values of "life-times" $t$ and decay
factors $k$ are chosen from some assumed distribution. The charges in the
vertices change in the simple possible way: $(0)\rightarrow (+) + (-)$,
$(+) \rightarrow (0) + (+)$ and $(-) \rightarrow (0) + (-)$. (see left
panel of Fig. \ref{Flow}). This code produces both the energy-momentum
distributions of produced particles and spatial distributions of their
production points (which show features of the truncated L\'evy
distributions). The example of the correlation function it leads to is
shown in Fig. \ref{C2M}.

\begin{figure}
\footnotesize \centerline{\begin{tabular}{@{}c@{}}
\includegraphics[height=1.0in]{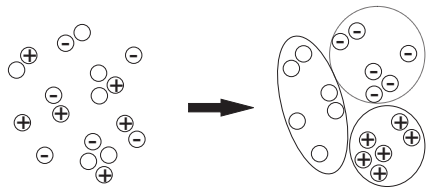} \\
\end{tabular}}
\caption[]{Charge reassignment method of generating effect of BEC in a
single event without changing its energy-momentum and spatio-temporal
characteristics \protect\cite{SP}.}\label{CasBec}
\end{figure}

CAS being classical scheme produces no BEC effect by itself. However, as
we have demonstrated in \cite{UWW,SP} the effect of BEC can be introduced
by special afterburner applicable to every separate event, but, contrary
to afterburners mentioned before, working without changing the number of
secondaries $N_{+/0/-}$ and without changing the energy-momentum and
spatio-temporal structure of this event. It is based on the observation
that effect of BEC can be visualized classically as phenomenon
originating due to correlations of some specific fluctuations present in
such stochastic systems as blob of the hadronizing matter
\cite{Optics,ZAJC,OMT}. It means that in phase space there occur bunches
of identical particles and BEC arises as {\it correlation of
fluctuations} effect. To get such effect in our case it is enough to
forget about the initial charge assignment in the event under
consideration (but keeping in mind the recorded multiplicities
$N_{+/0/-}$ and both energy-momenta and spatio-temporal positions of
particles emerging from NC code in event under consideration) and look
for bunches of identical particles, both in energy-momentum and
space-time. The procedure of selection of bunches of particles of the
same charge, is therefore the crucial point of such algorithm
\cite{UWW,SP}, cf. Fig. \ref{CasBec}. It generally consists in selecting
particles located nearby in the phase space, forming a kind of cell, and
endowing them with the same charge. In what follows we shall call such
cell {\it elementary emitting cell}, EEC. In this way we explore natural
fluctuations of spatio-temporal and energy-momentum characteristic of
produced particles resulting from CAS. Referring to \cite{UWW,SP} for
details let us only notice here that this methods works surprisingly well
and is very effective. Unfortunately it was not yet developed to fully
fledged MC code available for the common use but it allows to follow
changes made in the original CAS by imposing on it requirements or
reproducing also BEC patter. This is shown in Fig. \ref{Flow} where CAS
without BEC (left panel) is compared with CAS with BEC imposed. Whereas
$N_{+/0/-}$ and positions of all secondaries in phase space remain the
same the {\it charge flow pattern} changes considerably (see the right
panel of Fig. \ref{Flow}). BEC in this case enforces occurrence of {\it
multicharged vertices}, not present in the original CAS.

To summarize, the change in the original CAS required to observe BEC
pattern amounts to introduction of bunching of particles of the same
charge (in the form of EEC's). However, when done directly in the CAS it
would lead to great problems with the proper ending of cascade (without
producing spurious multicharged hadronic states, not observed in the
nature) and from this point of view, once we know what is the physics
behind BEC, the proposed afterburner algorithm occurs as a viable
numerical short cut solution to this problem. No such knowledge is,
however, provided for other afterburners used nowdays (although the idea
of bunching origin of BEC can be spotted in the literature, cf., for
example \cite{Bush}). The problem, which is clearly visible in the CAS
model, is not at all straightforward in other approaches. However, at
least in the string-type models of hadronization \cite{And}, one can
imagine that it could proceed through the formation of charged (instead
of neutral) color dipoles, i.e., by allowing formation of
multi(like)charged systems of opposite signs out of vacuum when breaking
the string. Because only a tiny fraction of such processes seems to be
enough in getting BEC in the case of CAS model, it would probably be
quite acceptable modification. It is worth to mention at this point that
there is also another possibility in such models, namely when strings are
nearby in the phase space one can imagine that production of given charge
with one string enhances emission of the same charge from the string
nearby - in this case one would have a kind of {\it stimulated emission}
discussed already in \cite{Pur,GN}.

\section{BEC from "within"}

The above observations, especially notion of EEC's, will be the
cornerstone of our new proposition. Let us remind that the idea of
bunching of particles as quantum statistical (QS) effect is not new and
has been used in the phenomenology of multiparticle production already
long time ago \cite{WJK,DeGS,JK,CMST,FT}. In connection with BEC it was
mentioned for the first time in \cite{Pur,GN} and later it formed a
cornerstone of the so called {\it clan model} of multiparticle
distributions $P(n)$ leading in natural way to their negative binomial
(NB) form observed in experiment \cite{NBD}. It was introduced in the
realm of BEC again in \cite{BSWW}, where the notion of EEC has been
introduced for the first time, and in \cite{OMT}.

\begin{figure}[h]
\footnotesize \centerline{\begin{tabular}{@{}c@{}}
\includegraphics[height=3.0in]{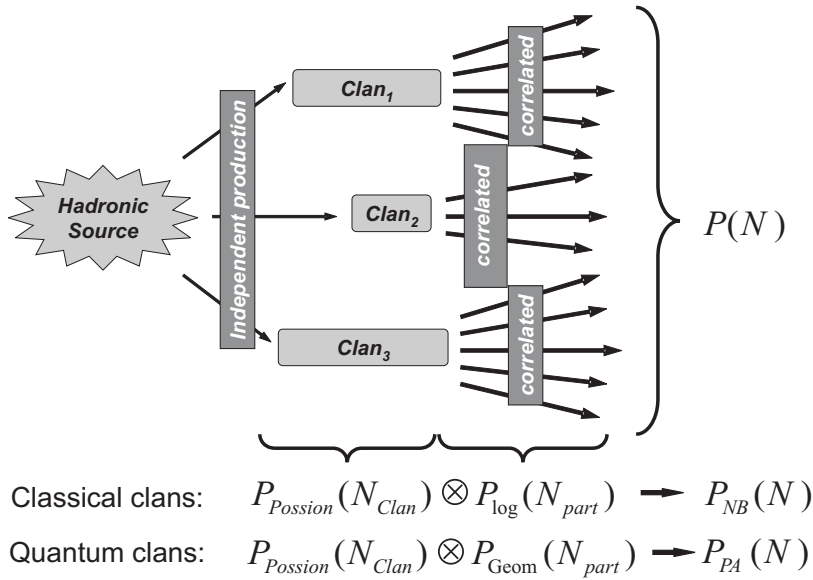} \\
\end{tabular}}
\caption[]{Schematic view of our algorithm, which leads to bunches of
particles ({\it clans}). Whereas in \cite{NBD} these clans could consist
of any particles distributed logarithmically in our case they consist of
particles of the same charge and (almost) the same energy and are
distributed geometrically to comply with their bosonic character. We are
therefore led to {\it Quantum Clan Model} \cite{QCM}.}\label{Q-clans}
\end{figure}

\begin{figure}
\footnotesize \centerline{\begin{tabular}{@{}cc@{}}
\includegraphics[height=1.6in]{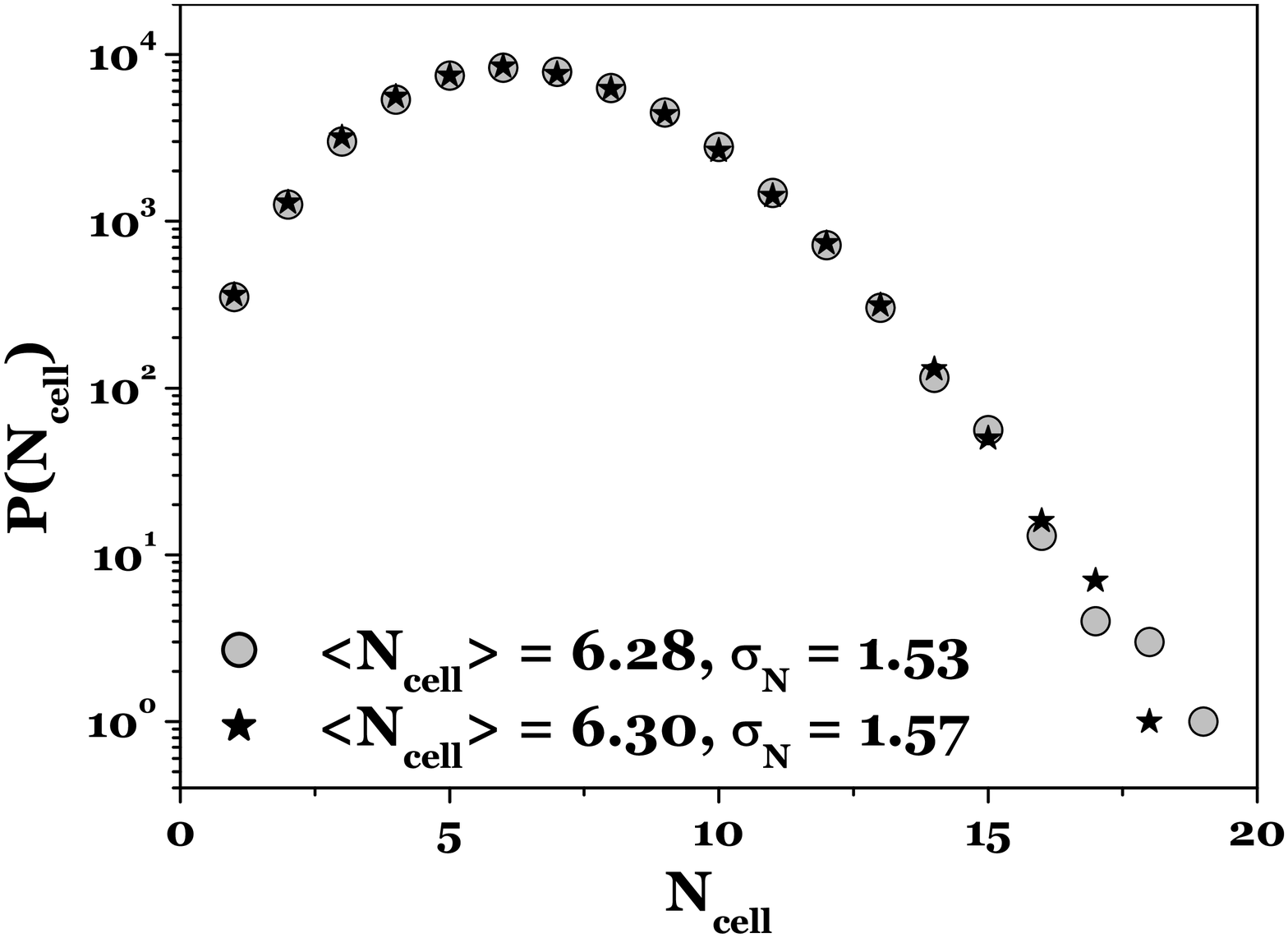} &
\includegraphics[height=1.6in]{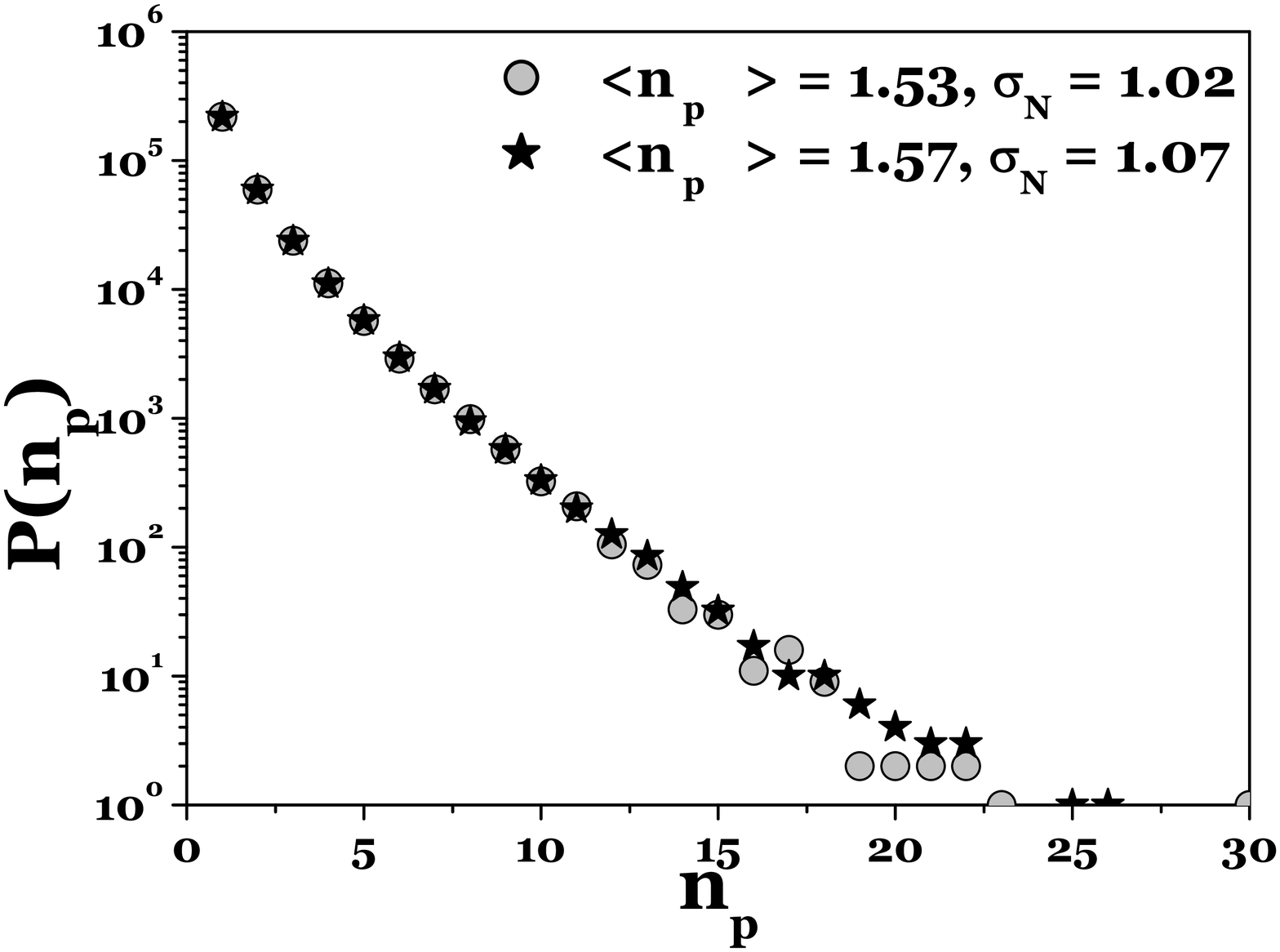}\\
\includegraphics[height=2.8in]{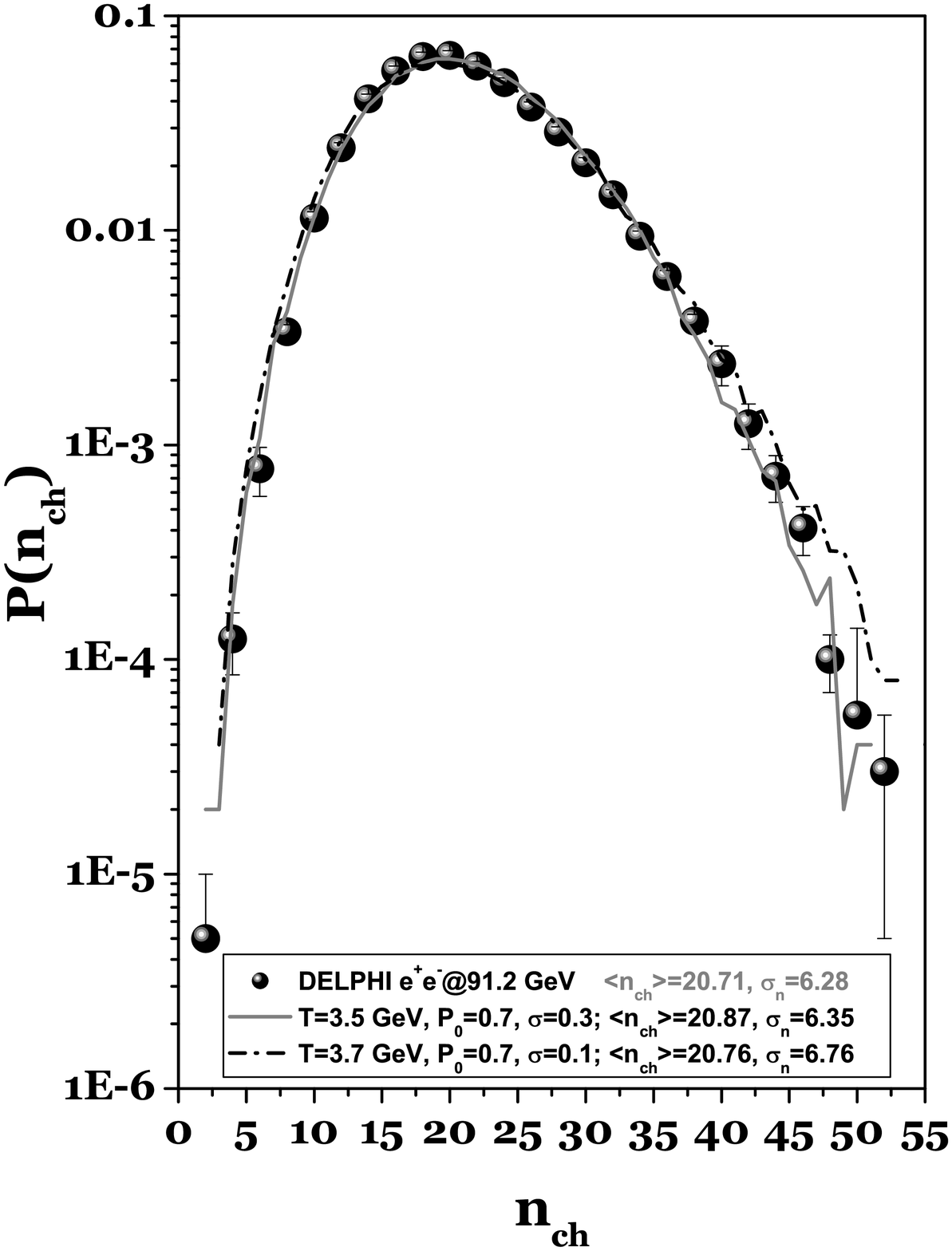} &
\includegraphics[height=2.8in]{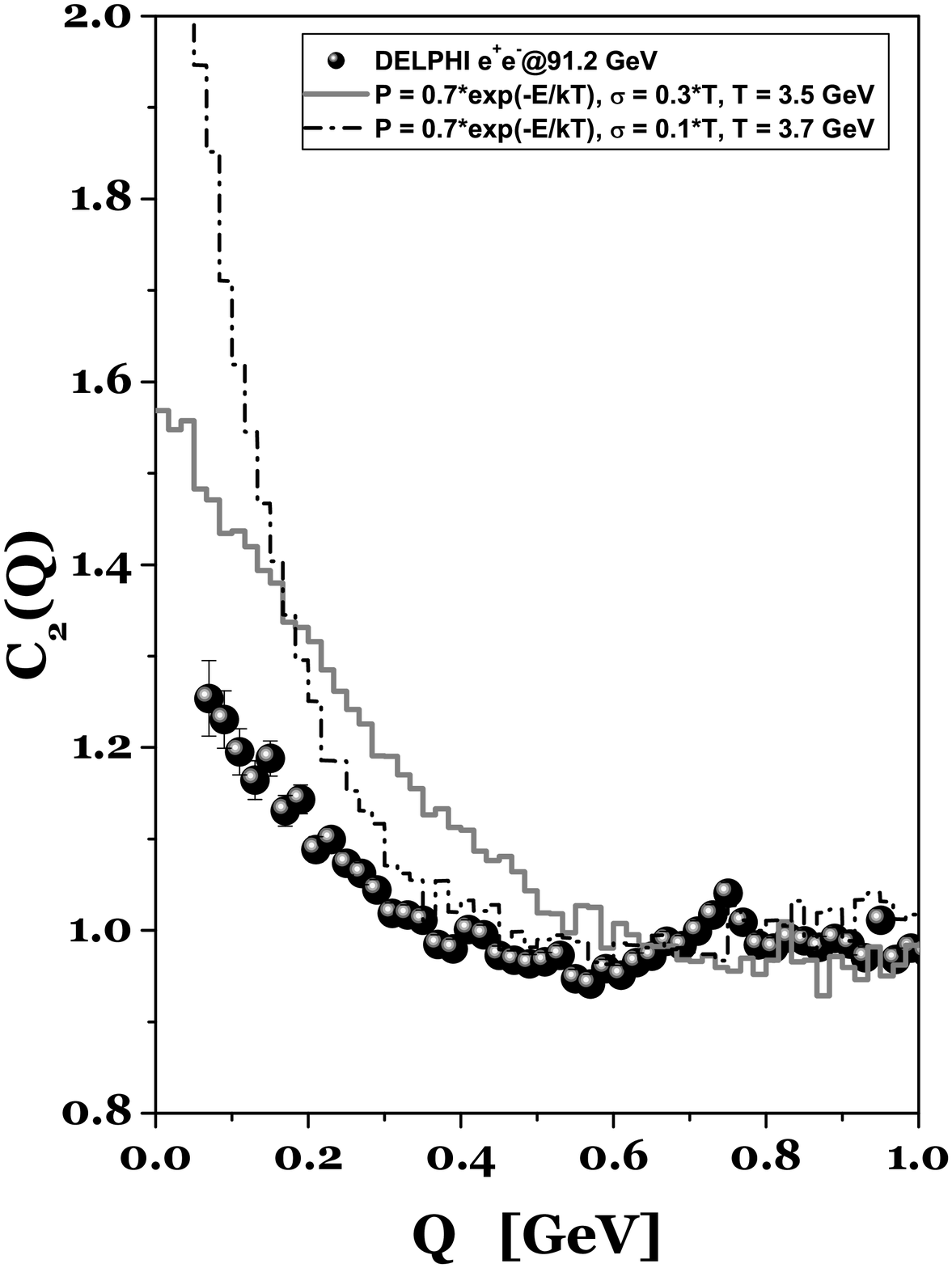}\\
\end{tabular}}
\caption[]{Upper panels: distribution of cells and particles in a given
cell. Lower-left panel: the corresponding summary $P(n)$ which is
convolution of both $P(n_{cell})$ and $P(n_p)$. Lower-right panel:
examples of the corresponding corresponding correlation functions
$C_2(Q)$ . Two sets of parameters were used. Data are from
\protect\cite{Data}.}\label{QC-examples}
\end{figure}

Because our motivation concerning viewing of BEC {\it from "within"}
comes basically from the work \cite{OMT} let us, for completeness,
outline shortly its basic points. It was the first MC code which intended
to provide as output distribution of particles containing already, among
its physical assumptions, the effect of BEC with no need for any
afterburner whatsoever. It deals with the problem of how to distribute in
phase space (actually in longitudinal phase space given by rapidity
variable) in a least biased way a given number of bosonic secondaries,
$\langle n\rangle =\langle n^{(+)}\rangle + \langle n^{(-)}\rangle +
\langle n^{(0)}\rangle$, $\langle n^{(+)}\rangle =\langle n^{(-)}\rangle
=\langle n^{(0)}\rangle$. Using information theory approach (cf.,
\cite{IT}) their single particle rapidity distribution was obtained in
the form of grand partition function with temperature $T$ and with
chemical potential $\mu$. To obtain effect of BEC the rapidity space was
divided into a number of {\it cells} of size $\delta y$ (which was fitted
parameter) each. Two very important observations are made there: $(i)$ -
whereas the very fact of existence of cells in rapidity space was enough
to obtain reasonably good multiparticle distributions, $P(n)$, (actually,
in the NB-like form) and $(ii)$ their size, $\delta y$, was crucial for
obtaining the characteristic form of the $2-$body BEC function
$C_2(Q=|p_i-p_j|)$ (peaked and greater than unity at $Q=0$ and then
decreasing in a characteristic way towards $C_2=1$ for large values of
$Q$) out of which one usually deduces the spatio-temporal characteristics
of the hadronization source \cite{BEC} (see \cite{OMT} for more details).
The message delivered was obvious: to get correlation function \cite{BEC}
\begin{equation}
C_2(Q=|p_1-p_2|) = \frac{d\sigma(p_1,p_2)}{d\sigma(p_1)\cdot
d\sigma(p_2)}
\end{equation}
peaked and greater than unity at $Q=0$ and then decreasing in a
characteristic way towards $C_2=1$ for large values of $Q$, one must have
particles located in cells in phase space which are of nonzero size. It
means therefore that from $C_2$ one gets not the size of the hadronizing
source, as it is frequently said, but only the size of the emitting cell,
$R\sim 1/Q$ \cite{Z} (in \cite{OMT} it is $R\sim 1/\delta y$).

It is worth to mention at this point that, as has been demonstrated in
\cite{Kozlov} in the quantum field theoretical formulation of BEC, the
requirement to get nonzero width od $C_2(Q)$ function corresponds
directly to the necessity of replacing delta functions of the type
$\delta(Q)$ in some commutator relations by a well defined, peaked at
$|Q|\rightarrow 0$ functions $f(Q)$ introducing in this way same
dimensional scale to be obtained from the fits to data. This fact was
known even before but without any phenomenological consequences
\cite{Zal}.

Let us now proceed to our proposition of looking on the problem of BEC.
As already mentioned, work \cite{OMT} has been our inspiration but we
would like to allow for dynamically defined EEC's and to assure
energy-momentum conservation. This was only approximate in \cite{OMT} and
rapidity cells there were fixed in size and were consecutively filled
with previously preselected number of particles. But already from our
previously described afterburner we have learned that EEC's can be of
different sizes (in fact, they even can overlap in phase-space)
\cite{SP}. What counts most is the fact that distribution of particles in
each EEC, $P(n_p)$, must follow geometrical distribution in order to get
the characteristic energy spectra for bosonic particles,
\begin{equation}
<n(E)> = \left[ \exp\left( E-\mu\right)/T -1 \right]^{-1}.\label{BEn}
\end{equation}

\begin{figure}
\footnotesize \centerline{\begin{tabular}{@{}cc@{}}
\includegraphics[height=1.8in]{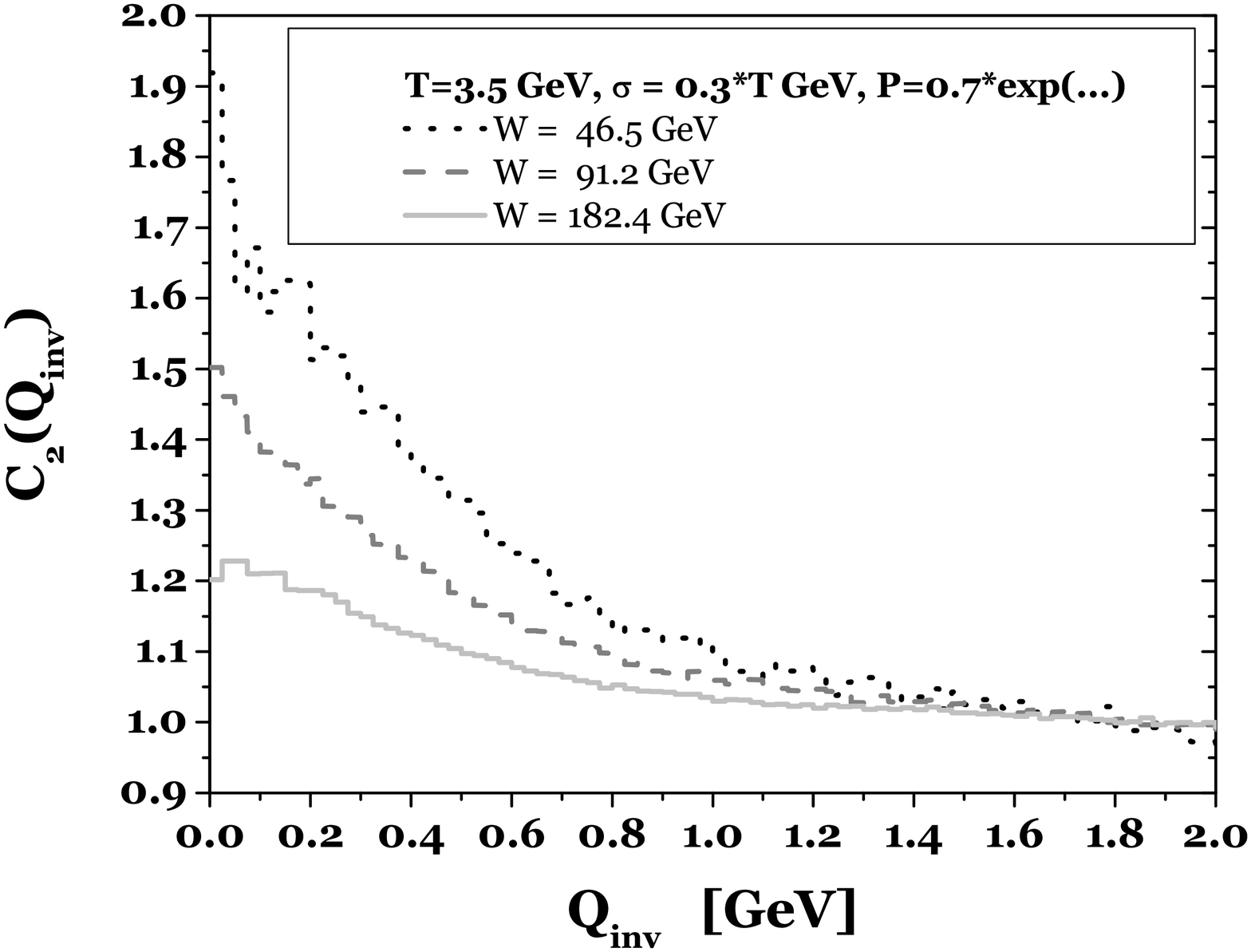} &
\includegraphics[height=1.8in]{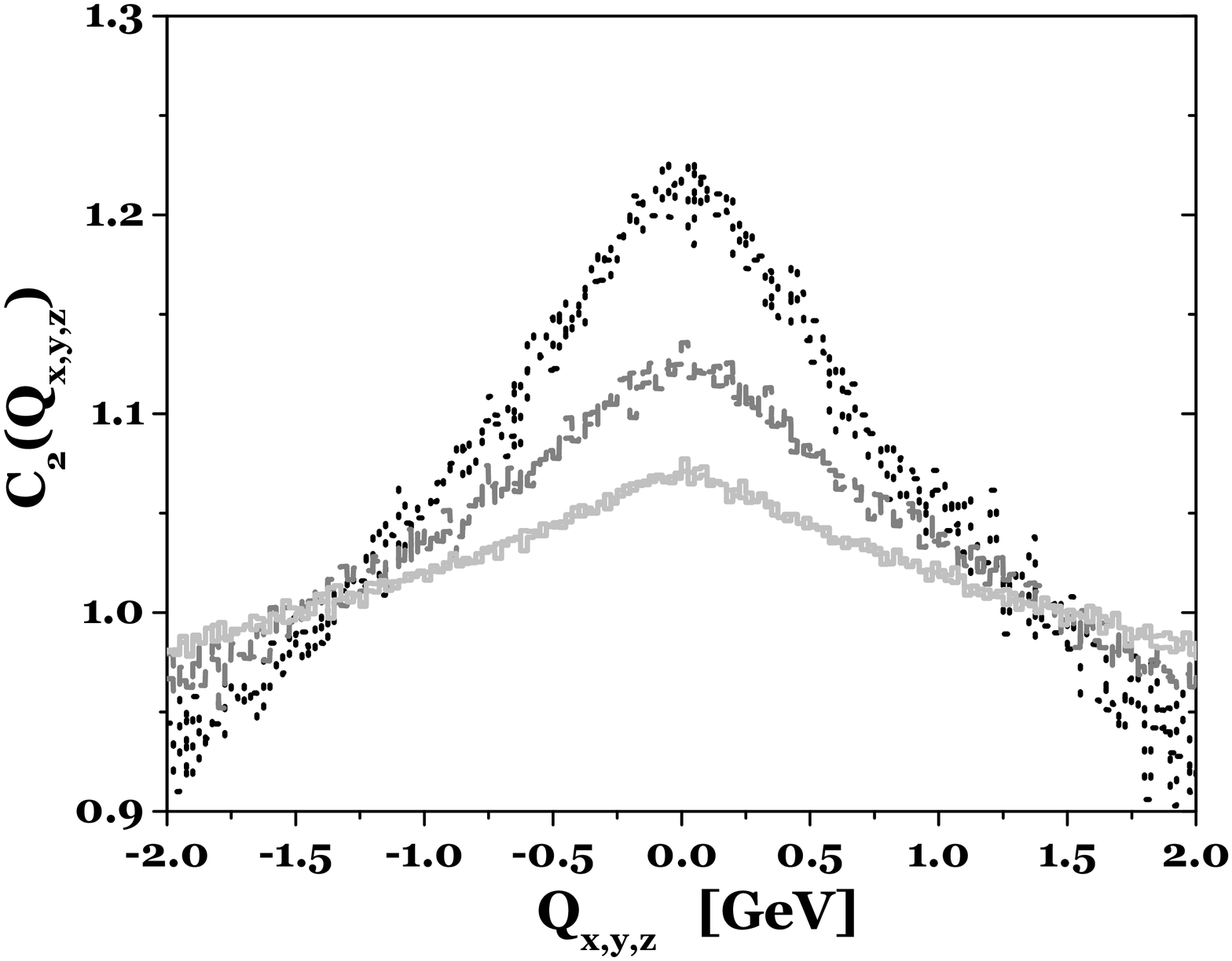}\\
\includegraphics[height=1.8in]{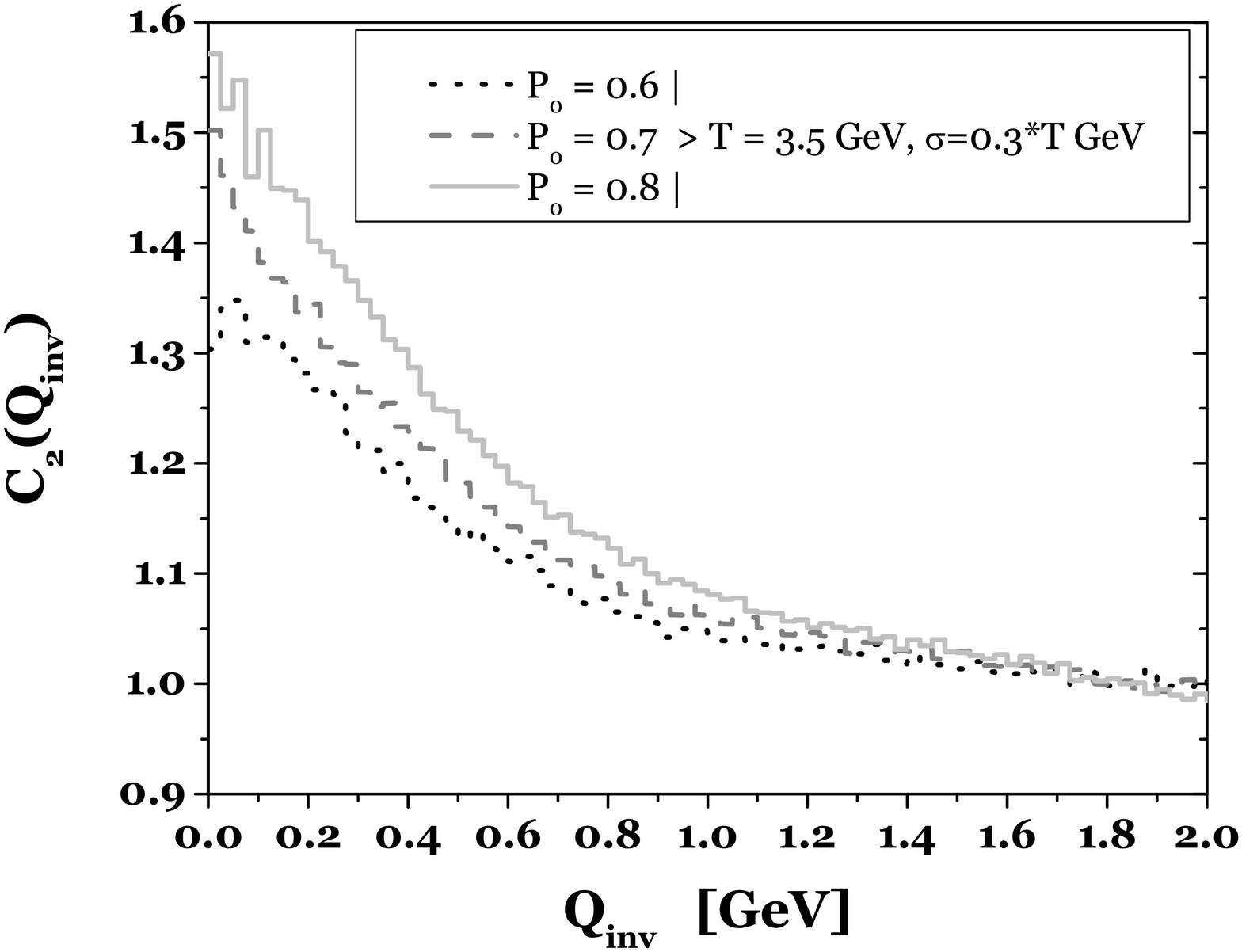} &
\includegraphics[height=1.8in]{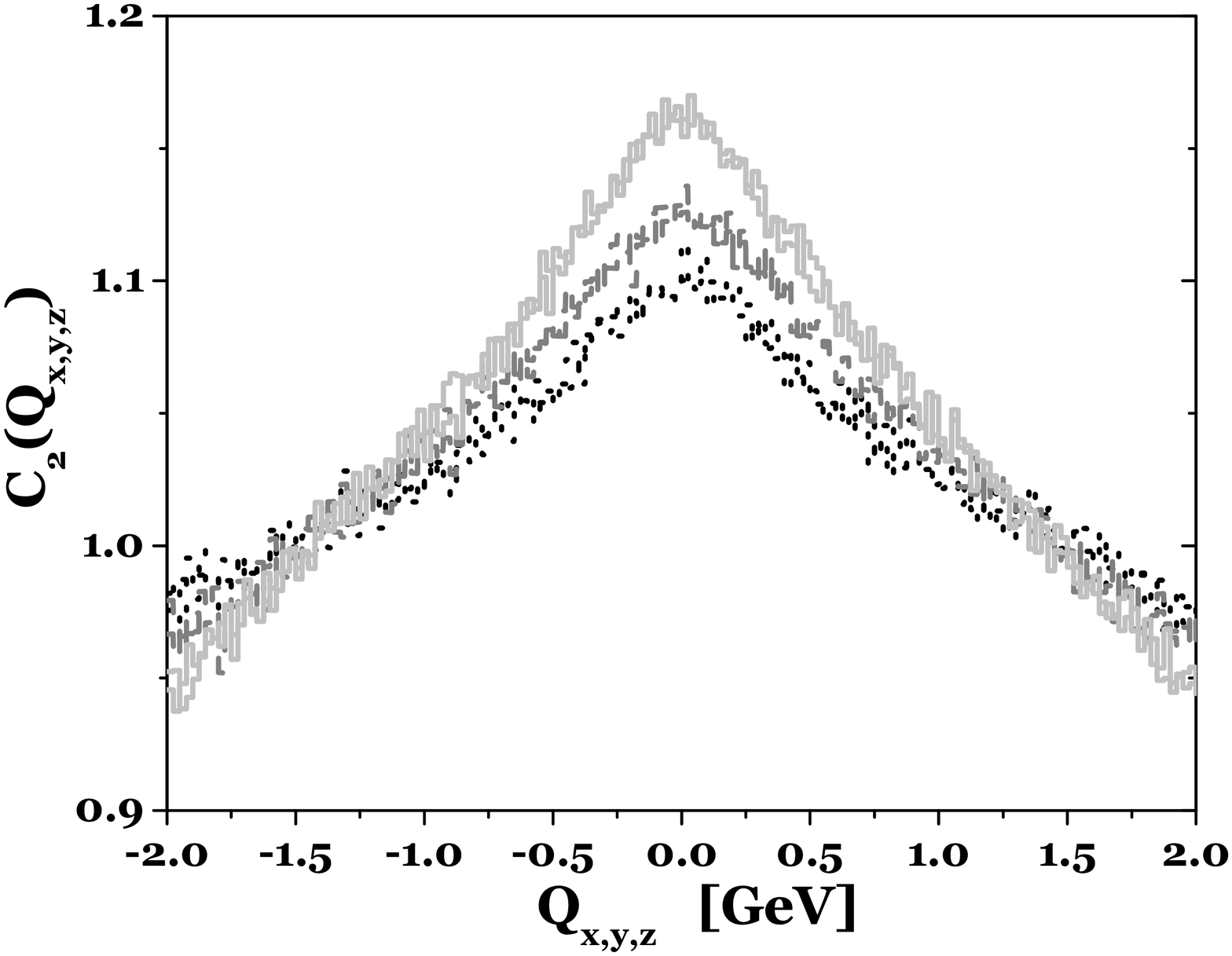}\\
\includegraphics[height=1.8in]{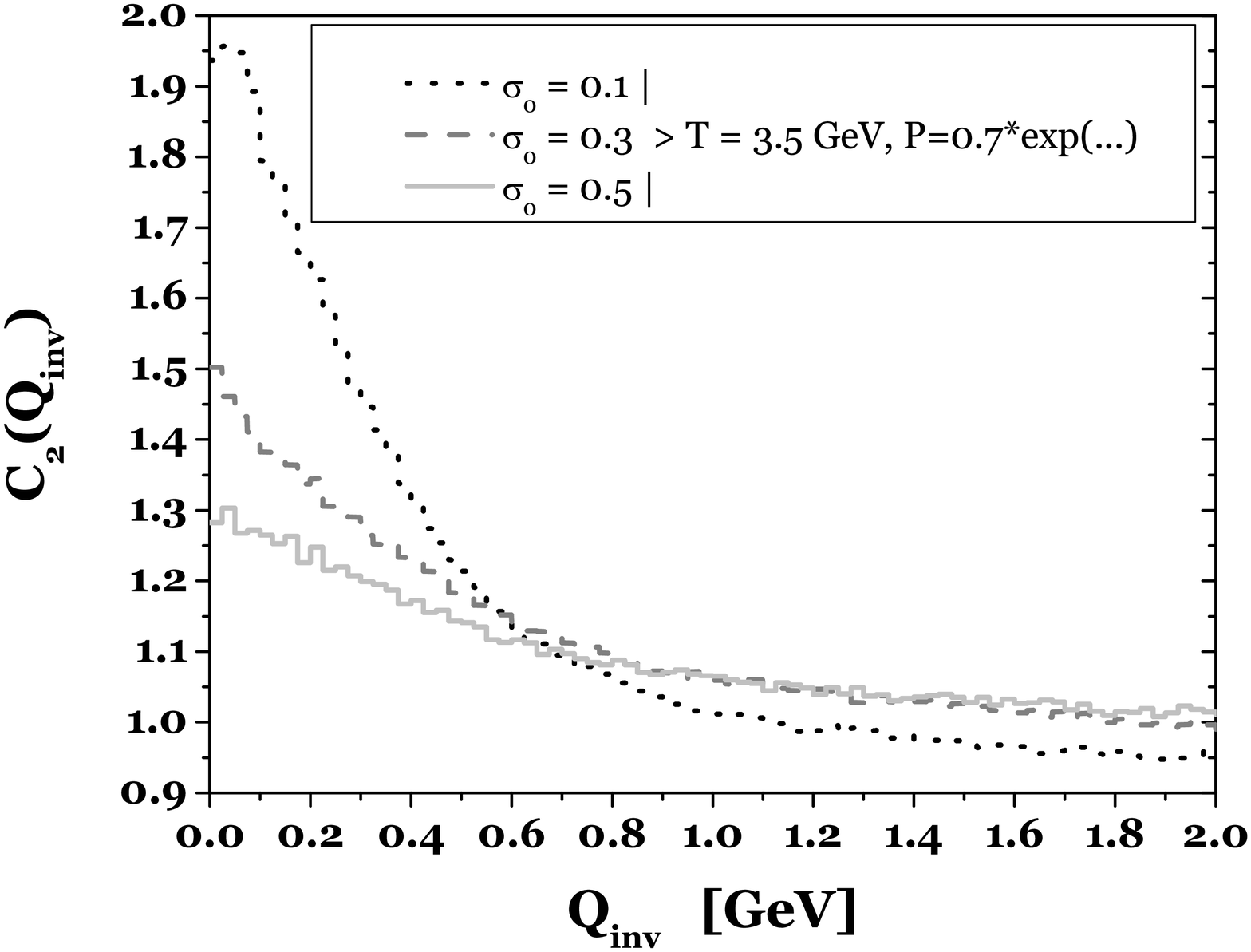} &
\includegraphics[height=1.8in]{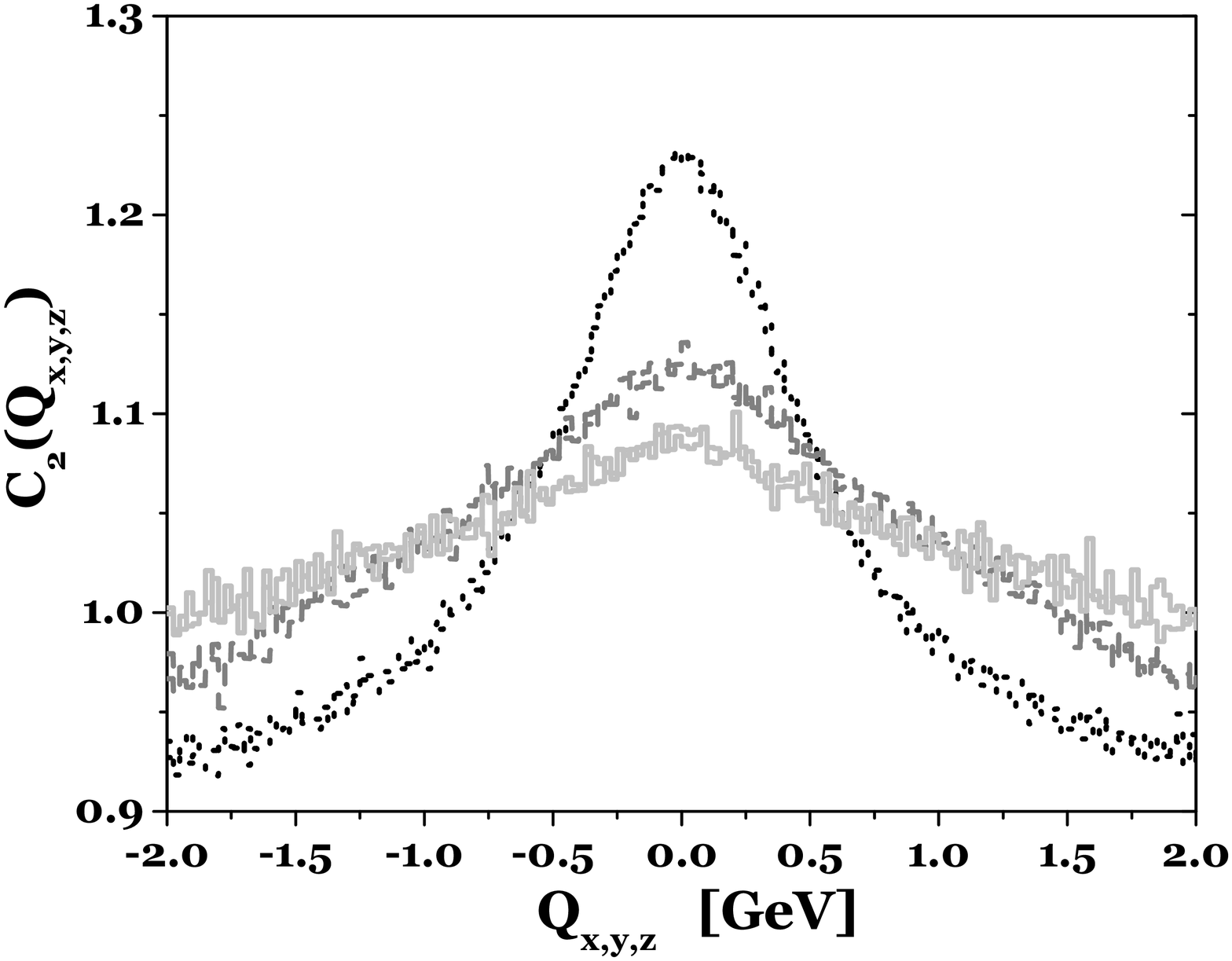}\\
\end{tabular}}
\caption[]{Example of results obtained for $C_2(Q_{inv})$ (left column)
and corresponding $C_2(Q_{x,y,z})$ (right column, for parameters used
here all $C_2(Q_{x,y,z})$ are the same). Calculations were performed
assuming spherical source $\rho(r)$ of radius $R=1$ fm (spontaneous decay
was assumed, therefore there is no time dependence) and spherically
symmetric distribution of $p_{x,y,z}$ components of momenta of
secondaries $p$. Energies were selected from $f(E) \sim \exp( -E/T)$
distribution. The changes investigated are - from top to bottom:
different energies of hadronizing sources, different values of parameter
$P_0$ and different spreads $\sigma$ of the energy in
EEC.}\label{Examplesofqxyz}
\end{figure}

To obtain such effect we proceed in the following way.
\begin{itemize}
\item[$(1)$] Using some (assumed) function $f(E)$ we select a particle of
energy $E^{(1)}_1$ and charge $Q^{(1)}$. The actual form of $f(E)$ should
reflect our {\it a priori} knowledge of the particular collision process
under consideration. In what follows we shall assume that $f(E) =
\exp\left( -E/T\right)$, with $T$ being parameter (playing in our example
the role of "temperature").

\item[$(2)$] Treating this particle as seed of the first EEC we add to
it, with probability $P(E)=P_0\cdot\exp\left( -E/T\right)$, until the
first failure, other particles of the same charge $Q^{(1)}$. Such
procedure results in desired geometrical (or Bose-Einstein) distribution
of particles in the cell, i.e., in
\begin{equation}
\langle n_p\rangle = P(E)/([1+P(E)], \label{nBE}
\end{equation}
accounting for their bosonic character. Together with exponential factor
in probability $P(E)$, it assures that occupancy number of state with
given energy will eventually follow characteristic bosonic form as given
by eq. (\ref{BEn}) (here $P_0$ is another parameter playing the role of
"chemical potential" $\mu = T\cdot \ln P_0$). As result $C_2(Q)>1$ but
only {\it at one point}, namely for $Q=0$.

\item[$(3)$] Because process of emission of particles in a given EEC has
finite duration, the resultant energies must be spread around the energy
of the particle defining given EEC by some amount $\sigma$ (which is
another free parameter). It automatically leads to the experimentally
observed widths of $C_2(Q)$.

\end{itemize}

Points $(1)$-$(3)$ are repeated until all energy is used. Every event is
then corrected for energy-momentum conservation caused by the selection
procedure adopted and condition $N^{(+)}=N^{(-)}$ is imposed as well.

As result in each event we get a number of EEC with particles of the same
charge and (almost) the same energy, i.e., picture closely resembling
classical {\it clans} of \cite{NBD} (with no effects of statistics
imposed, see Fig. \ref{Q-clans}). Our clans (i.e., EECs containing
identical bosonic particles subjected to quantum statistics and therefore
named {\it quantum clans}) are distributed in the same way as the
particles forming the seeds for EEC, i.e., according to Poisson
distribution. With particles in each clan distributed according to
geometrical distribution they lead therefore to the overall distribution
being of the so called P\`olya-Aeppli type \cite{JF}. This distribution
strongly resembles the Negative Binomial distributions obtained in the
classical clan model \cite{NBD} where particles in each clans were
assumed to follow {\it logarithmic} distribution instead (with
differences occurring for small multiplicities \cite{DWH}). The first
preliminary results presented in Fig. \ref{QC-examples} are quite
encouraging (especially when one remembers that so far effects of
resonances and all kind of final state interactions to which $C_2$ is
sensitive were neglected here).

The main outcome so far is strong suggestion that EEC's are among the
possible explanations of the BEC effect, in which case BEC provide us
mainly with their characteristics, not with the characteristics of the
whole hadronizing source. So far our method applies only to
one-dimensional example of $C_2(Q)$. This is because process of formation
of EEC's proposed here gives us energies of all particles in a given cell
and therefore also $p_i =|\vec{p}_i|$, but says us nothing about their
angular distributions (i.e., about their components $p_{ix,iy,iz}$). To
extend it to three dimensional case of $C_2(Q_{x,y,z})$ one has to
somehow build cells also in $(p_x,p_y,p_z)$ components of the momenta of
particles forming EEC. This can be done in many ways. Here we present as
example approach using pairwise symmetrization of the wave functions of
every $i>1$  particle in a given EEC with the first particle ($i=1$)
defining it. As result one gets for each such pair (in the plane wave
approximation \cite{BEC} and assuming instantaneous hadronization) the
known $1+\cos\left[ \left(\vec{p}_1 - \vec{p}_i\right)\cdot
\left(\vec{r}_1 - \vec{r}_i\right)\right]$ term which connects the
spatial extension of the hadronizing source $\rho(r)$ with is momentum
space characteristics obtained before. In this way the assumed shape of
$\rho(r)$ translates into the respective cells in momentum space.
Preliminary results of this procedure are presented in Fig.
\ref{Examplesofqxyz}.

\section{Summary}

We propose new numerical method of accounting for BEC phenomenon from the
very beginning of the modelling process. Once the expected energy
spectrum $f(E)$ of produced particles is chosen one constructs EEC's in
energy. We regard this method as very promising but we are aware of the
fact that our proposition is still far from being complete. To start with
one should allow for time depending emission by including $\delta E\cdot
\delta t$ term in the $\cos(\dots)$ above. The other is the problem of
Coulomb and other final state interactions. Their inclusion is possible
by using some distorted wave function instead of the plane waves used
here. Finally, so far only two particle symmetrization effects have been
accounted for: in a given EEC all particles are symmetrized with the
particle number $1$ being its seed, they are not symmetrized between
themselves. To account for this one would have to add other terms in
addition to the $\cos(\dots)$ used above - this, however, would result in
dramatic increase of the calculational time.

We shall close with remark that there are also attempts in the literature
to model numerically BE condensation effect \cite{KR,KKRP} (or to use
notion of BE condensation in other branches of science as well
\cite{ClassBE,St}) using ideas of bunching of some quantities in the
respective phase spaces.

\acknowledgements GW is grateful for the support and warm hospitality
extended to him by organizers of the International Conference NEW TRENDS
IN HIGH-ENERGY PHYSICS (experiment, phenomenology, theory), Yalta,
Crimea, Ukraine, September 10-17, 2005, conference. Partial support of
the Polish State Committee for Scientific Research (KBN) (grant
621/E-78/SPUB/CERN/P-03/DZ4/99 (GW)) is acknowledged.


\end{document}